\documentstyle[amssym,11pt,aaspp4,flushrt]{article} 

\slugcomment{Accepted for publication in Astrophysical Journal}

\lefthead{Chadwick et al.}

\righthead{Gamma Rays from Centaurus X-3}

\begin{document}
                          
\title{Centaurus X-3 --- A Source of High Energy Gamma Rays} 

\author{P.~M.~Chadwick, M.~R.~Dickinson, N.~A.~Dipper, T.~R.~Kendall,
T.~J.~L.~McComb, K.~J.~Orford, J.~L.~Osborne, S.~M.~Rayner,
I.~D.~Roberts, S.~E.~Shaw and K.~E.~Turver}

\affil{Department of Physics, Rochester Building,
Science Laboratories, University of Durham, Durham, DH1~3LE, U. K.}

\begin{abstract} 

Centaurus X-3 is a well-studied high-mass accreting X-ray binary and a
variable source of high energy gamma rays with energies from 100 MeV to
1 TeV. Previous results have suggested that the origin of the gamma rays
is not the immediate vicinity of the neutron star but is sited in the
accretion disc, perhaps in an accretion wake. The Durham Mark 6
gamma ray telescope has been used to measure the gamma ray flux from
\mbox{Centaurus X-3} with much higher sensitivity than previous
ground-based measurements. The flux above $\sim 400$ GeV was measured to
be $2 \pm 0.3 \times 10^{-11}~\mathrm{cm}^{-2}~\mathrm{s}^{-1}$ and
appears constant over a period of 2 -- 3 months. In 10 hours of
observations there is no evidence for periodicity in the detected gamma
rays at the X-ray spin period either from a site in the region of the
neutron star, or from any other potential site in the orbit.

\end{abstract}


\keywords{stars: individual (Cen X-3) --- gamma rays: observations}


\section{Introduction}

The high-mass X-ray binary \mbox{Centaurus X-3} was the first binary
from which coherent X-ray pulses were detected (\cite{kn:giacconi}). It
contains a 4.8 s pulsar in a 2.1 day orbit around an O-type supergiant,
V779 Cen. The pulsar period has been shortening since discovery; this is
attributed to accretion of matter by the neutron star from the more
rapidly rotating inner edge of its accretion disc. The long-term period
history was monitored by {\em GINGA} and shown to be a linear variation
with a small added sinusoidal term of about 9 year period
(\cite{kn:tsunemi}). However, this long-term periodicity has not been
observed in recent data from BATSE (\cite{kn:bildsten}). Large
short-term variations in both X-ray amplitude and instantaneous period
derivative are seen with no apparent correlations (\cite{kn:finger};
\cite{kn:tsunemi}), suggesting that the X-ray luminosity observed is
often only a small fraction of the accretion power.

Very high energy gamma rays (TeV range) were detected by us using an
earlier, less sensitive, atmospheric \v{C}erenkov telescope
(\cite{kn:alberto}; \cite{kn:brazier}) and by another group also using
\v{C}erenkov telescopes (\cite{kn:raub}). These observations, which were
based on searches for pulsation, showed evidence for sporadic strong
emission at a preferred orbital phase and for pulsations at, or in some
cases near to, the pulsar period (\cite{kn:lamb}). We suggested that the
emission may not be from a site close to the neutron star
(\cite{kn:bowden}). A recent paper gives evidence, from a re-analysis of
archival data, for gamma ray emission from a site in the the accretion
disc trailing the neutron star by $70^{\circ}$ (\cite{kn:raubsmit}).

A GeV gamma ray outburst has been detected by the EGRET telescope on
board {\em CGRO} (\cite{kn:vest}) at 100 MeV with evidence for pulsation
at the pulsar period with an average luminosity $\sim 5 \times
10^{36}~\mathrm{erg}~\mathrm{s}^{-1}$, a significant fraction of the
total accretion luminosity which can vary from $\sim 4 \times
10^{37}~\mathrm{erg}~\mathrm{s}^{-1}$ (\cite{kn:finger}) to $\sim 1.2
\times 10^{38}~\mathrm{erg}~\mathrm{s}^{-1}$ (\cite{kn:white}). 

We report here our first observations with an imaging telescope of much
higher sensistivity and a lower energy threshold than previously
employed on this source.


\section{The Mark 6 Telescope}

The Durham Mark 6 atmospheric \v{C}erenkov telescope has been described
in detail elsewhere (\cite{kn:armstrong}). It comprises three parabolic
flux collectors of 7 m diameter and aperture f/1.0, mounted on a single
alt-azimuth platform. At the focus of the central dish is a
thermostatically controlled camera of $3^{\circ}$ diameter comprising
109 photomultipliers in a close-packed array with hexagonal reflective
cones to reduce light lost between photocathodes. The other two
identical parabolic collectors each have a temperature-controlled
triggering camera containing 19 larger photomultipliers which cover an
area similar to the central imaging camera. Events are recorded when
there are responses in all three cameras which correlate in position at
the foci (within $0.5^{\circ}$ and in time to within 10 ns). This type
of triggering gives this telescope a unique ability to detect low energy
gamma rays from the ground with immunity from triggering by local muons,
which are a significant limiting factor for low energy detection in
single-mirror telescopes. The presence of independent noise in each of
the triggered channels of the Mark 6 telescope results in a threshold
which is not sharply defined. The probability of gamma ray detection
reduces slowly from $\sim 40\%$ at an energy of 250 GeV to $\sim 1\%$ at
an energy of 100 GeV.

Comprehensive real-time calibration of the photomultipliers and
digitizing electronics is performed by triggering the telescope at
random times using (a) a nitrogen laser/plastic scintillator/optical
fibre guide/opal diffuser system to simulate \v{C}erenkov flashes and
enable flat-fielding, and (b) pulses injected after the coincidence
stage to provide random samples of the background noise. The latter
enables a correct pedestal or zero to be assigned to each fast ADC to an
accuracy of $\pm 0.1$ digital counts. The laser trigger enables the gain
of each photomultiplier to be measured relative to the others, to an
accuracy of about 1\% for each 15-minute data segment. Each type of
random trigger is generated at an average rate of 50 per minute.

Simultaneous and continuous monitoring of the atmospheric clarity is
carried out using an infra-red radiometer monitoring the source region
and an axial optical CCD camera which enables the position and magnitude
of guide stars to be measured. The digital data record for each event
contains: (1) the digitized charge from each photomultiplier, (2) the
time of each event, to an accuracy of 1 $\mu$s, using a GPS-moderated
rubidium oscillator, and (3) the telescope pointing position to 1 arc
minute. Every minute a record is made of the output of a comprehensive
weather station, and of the temperatures inside the photomultiplier
packages and the electronics, all important system voltages, and the
anode currents and trigger rates of all photomultipliers.

\section{Observations}

\mbox{Centaurus X-3} was observed using the Durham Mark 6 \v{C}erenkov
telescope for typically 3 hours per night, from 0900 UTC on 1997 March
1, 3 and 4 (corresponding to orbital phases (relative to X-ray eclipse)
of 0.78, 0.74 and 0.22, respectively) and from 1030 UTC on 1997 June 1,
2, 4, 5 and 7 (corresponding to orbital phases 0.89, 0.37, 0.33, 0.81
and 0.77, respectively). The majority of our observations were targeted
to be near orbital phase 0.8 with respect to X-ray eclipse at which TeV
emission had been detected earlier (\cite{kn:brazierb}). Data were taken
in 15-minute segments alternately on- and off-source. The total cosmic
ray background counting rate of the telescope was $\sim 400$ per minute.
The on-source segments were taken with \mbox{Centaurus X-3} near, but
not exactly at the center of the camera. The source was usually
$0.1^{\circ}-0.2^{\circ}$ from the camera center. Off-source data were
taken by observing a section of sky at the same declination as
\mbox{Centaurus X-3}, but separated by 15 minutes in right ascension.
All data segments started so that the off-source and on-source segments
possessed identical azimuth/zenith profiles. At the declination of
\mbox{Centaurus X-3} ($-60.6^{\circ}$), the center of the camera was
pointed during the off-source segments at a location which was
$1.84^{\circ}$ from the source position.

Data segments were accepted for analysis only if the sky was clear and
stable and the total counting rates in each on/off-source pair were
consistent at the $2.5~\sigma$ level. After these requirements were
satisfied a data set containing a total of 20 hours, equally divided
between on-source and off-source observations, was analysed further.

\section{Analysis}

Routine reduction and analysis of accepted data comprises:

\begin{enumerate}
 
\item calibration of gains and pedestals of all 147 photomultipliers and
ADC electronics within a 15-minute segment, using the embedded laser and
false coincidence events, 

\item software padding of the data (\cite{kn:fegan}) to equalize the
on-source and off-source photomultiplier noise, 

\item identification of the precise location of the source in the
camera's field of view for each event, using the axial CCD camera, 

\item a calculation of the spatial moments of the shower image relative
to the source position, for each event, 

\item a rejection of events containing images which would be unlikely to
be produced by gamma rays.

\end{enumerate}

The image parameters used for selection of gamma rays are found using
the established `image and border' technique (\cite{kn:fegan}). These
parameters are: {\em BRIGHTNESS}, the total number of digital counts in
the image, {\em DISTANCE} of the image centroid from the source
position, {\em ECCENTRICITY} of the image (width/length), {\em ALPHA},
the angle by which the image's long axis misses the source and {\em WIDTH},
the RMS spread of the image along the minor axis.

In addition to these parameters, we are able to exploit another property
of gamma ray shower images: the relative freedom from fluctuations in
the images from gamma ray showers in separated detectors. We had
demonstrated this in the enhancement of the gamma ray signal from AE
Aquarii using two well-separated (100 m) telescopes
(\cite{kn:chadwick}). In the Mark 6 detector the left and right
triggering cameras are 15 m apart and can accurately measure similar
parameters. These parameters depend on only the lower moments of the
images: brightness and centroid position in the focal plane. Of those
developed in \cite{kn:chadwick}, the parameter $D_{\mathrm{miss}}$ is
least affected by the relatively small separation of the detectors; we
denote the single-telescope analog of this parameter by
$D_{\mathrm{dist}}$. This measures the angular separation in the focal
planes of the centroids of the two images. The sensitivity of this
measure to cascade development has been verified by Monte Carlo
simulations of showers and by observing the variation in the
distribution of the parameter with zenith angle, using off-source data.
The expected variation of the parallax due to shower development changes
with zenith angle is clearly seen. Observations of gamma rays from PSR
B1706-44 confirm the value of the Q-factor\footnote{defined as
\[\frac{\rm{fraction~of~gamma~
rays~retained}}{\sqrt{\rm{fraction~of~background~retained}}}\]} for this
technique to be $\sim 1.4$, in agreement with simulation results
(\cite{kn:chad_durban}; {\cite{kn:chad_astropart}).

\section{Results}

Before background rejection, events are removed from the data which have
images of very low brightness or whose centroids fall very close to the
camera edge, more than $1.1^{\circ}$ from the center of the camera.
After these have been removed there are 64713 events in the on-source
data set and 65171 in the off, a source deficit of $1.3~\sigma$. The
parameter limits were set as follows:

\begin{enumerate}
 
\item $800 > {\em BRIGHTNESS} > 20000$ total digital counts
(corresponding to $\gtrsim 400$ GeV), 

\item $0.3 < {\em ECCENTRICITY} < 0.8$, required to
enable {\em ALPHA} to be estimated without large errors, 

\item $0.35^{\circ} < {\em DISTANCE} < 0.75^{\circ}$, required to allow
gamma ray shower images to be elongated (lower limit) and to reduce
effects due to the edge of the camera (upper limit),
 
\item $D_{\mathrm{dist}} < 0.07^{\circ}$,
 
\item ${\em WIDTH} < 0.28^{\circ}$.

\end{enumerate}

After these selections, there were 4429 events on-source and 4087
off-source, an on-source excess of significance $3.7~\sigma$. There has
been no normalisation of any of the on- and off-source parameter
distributions. The distribution of {\em ALPHA} shown in
Figure~\ref{fig:alpha} gives clear evidence of the excess being at small
values of {\em ALPHA} were the excess is expected from a true gamma ray
source (see e.g. \cite{kn:fegan}). For {\em ALPHA} less than $30^{\circ}$, the
on-source and off-source data sets contain 1546 and 1208 events
respectively, a \mbox{$6.4~\sigma$} excess: $338 \pm 52$ events. 

\begin{figure}[tb] 

\epsscale{0.5}
\plotone{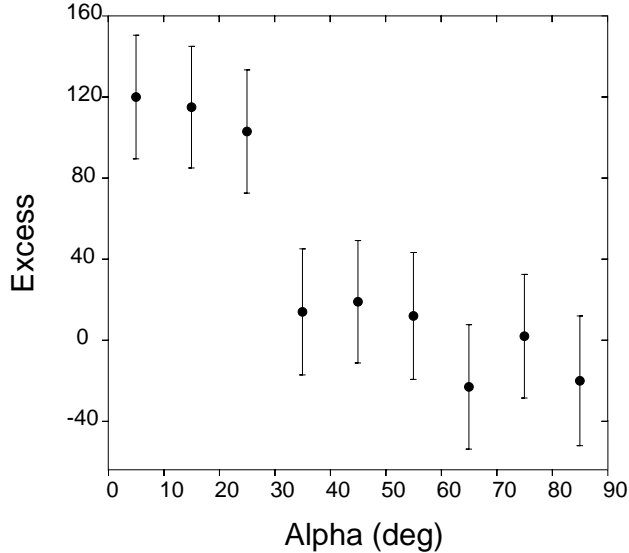}

\caption{The difference of the distributions in {\em ALPHA} for
on-source data and off-source data.}

\label{fig:alpha} 
\end{figure} 

A false-source analysis has been performed by re-analysing the on- and
off-source data using a matrix of trial source positions, in celestial
coordinates. At each trial source position in the matrix, {\em DISTANCE}
and {\em ALPHA} are recalculated for each event and a final selection of
events is made with the {\em DISTANCE} selection as above and {\it
ALPHA} $< 30^{\circ}$. The excess events for each trial source position
are shown in Figure~\ref{fig:falsesource}. The central point in the
matrix is the position of \mbox{Centaurus X-3} and is not the center of
the camera, which is typically off-source by $0.1^{\circ} -
0.2^{\circ}$. The width of the central peak is determined by the maximum
value of {\em ALPHA} in the selection and the maximum and minimum values
selected for {\it DISTANCE}. A very strong source would allow tighter
selections and a narrower peak.

\begin{figure}[tb]

\epsscale{0.5}

\plotone{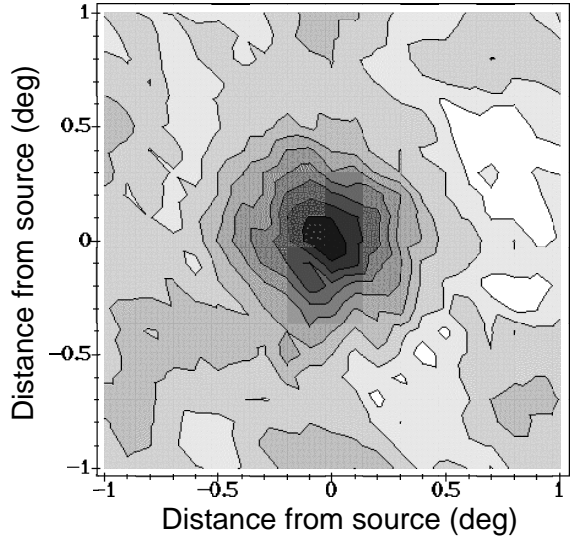}

\caption{False source analysis: excess events with {\em ALPHA} less than
$30^{\circ}$. The contours are spaced at $0.6~\sigma$ intervals, with
the grey scale being such that black indicates a probability of $> 6
\sigma$ for a gamma ray originating from that direction.}

\label{fig:falsesource}
\end{figure} 


\subsection{Flux}

The 338 excess gamma rays were detected in 10 hours of on-source
observation. The selection procedure retained $\sim$ 50\% of the
original gamma ray events. The collecting area has been estimated using
Monte-Carlo simulations (\cite{kn:armstrong}) and is $\sim
10^{5}\:\mathrm{m}^{2}$ at the energies and zenith angles at which Cen
X-3 was observed. Using this estimate and assuming a differential flux
index of $-1.8$, the gamma ray flux above 400 GeV was $(2 \pm 0.3)
\times 10^{-11}~\mathrm{cm}^{-2}~\mathrm{s}^{-1}$. The flux would be
$\sim 10\%$ higher for an assumed spectral index of $-2.6$. The main
uncertainty is not in the collecting area, which is dominated by
geometric effects and the height of origin of the electromagnetic
shower, but by the variation of triggering probability with energy. This
may be estimated from simulations and checked by observations, but still
may be in error by $\sim 50\%$. Our result is shown in
Figure~\ref{fig:spectrum} where it is compared with previous results,
including the EGRET spectrum obtained from a recently reported burst of
GeV gamma rays (\cite{kn:vest}). The burst occured during the week
starting 1994 October 18 and had an integral flux above 0.1 GeV of $9.2
\pm 2.3 \times 10^{-7}~\mathrm{cm}^{-2}~\mathrm{s}^{-1}$ with a best fit
differential index of $-1.81 \pm 0.37$ in the range 0.1 to 10 GeV.
Figure~\ref{fig:spectrum} also shows a 2 $\sigma$ upper limit form an
earlier EGRET exposure in 1994 January.

\begin{figure}[tb]
\epsscale{0.5}
\plotone{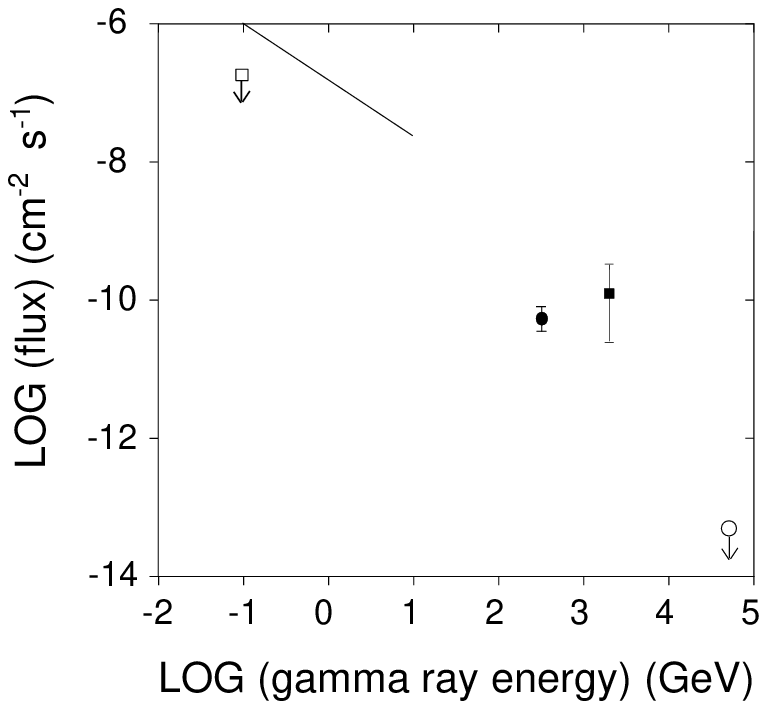}

\caption{Gamma ray integral number flux spectrum. The solid line is the
EGRET 1994 October burst (\cite{kn:vest}). Key: $\bullet$ -- this work,
$\square$ -- upper limit from January 1994 EGRET data (\cite{kn:vest}),
$\blacksquare$ -- \cite{kn:north}, $\circ$ -- \cite{kn:yoshi}.}

\label{fig:spectrum}
\end{figure} 

Our data have been examined for short-term variations. The detected rate
of gamma rays per hour is shown for the observations on eight nights in
Figure~\ref{fig:nightly}. The night-by-night detected gamma ray rates
show no significant deviation from constancy ($\chi^2 = 9$ for 8 degrees
of freedom). We have also tested to see if there is any significant
difference between the detected count rates for observations clustered
around phase 0.25 and phase 0.75. We find that there is a $1.7 \sigma$
difference to which we ascribe no significance.

\begin{figure}[tb]
\epsscale{0.5}

\plotone{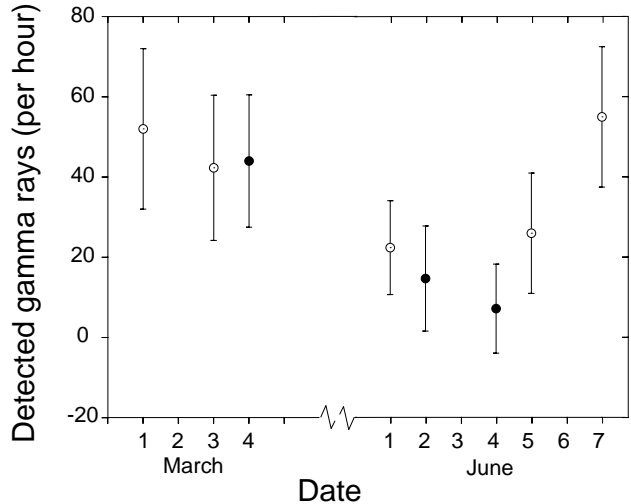}

\caption{Rate of detected gamma rays per hour for each night's
observation. Data taken at orbital phases with respect to mid X-ray
eclipse: $\circ$ $\;\sim\,$0.25, $\bullet$ $\;\sim\,$0.75.}
\label{fig:nightly} 

\end{figure} 

\subsection{Periodicity}

A data set was made from on-source events selected to have {\em ALPHA}
$< 30^{\circ}$ and the exact event times were reduced to the Solar
System Barycenter using the \mbox{DE200} ephemeris and then to the
position of the neutron star using the current orbital ephemeris
employed by \mbox{BATSE} (\cite {kn:bildsten}).

Data from individual nights have been searched for periodicity using the
Rayleigh test. The period search range was over one independent Fourier
interval about the contemporary X-ray period taken from the BATSE
results\footnote{Available on the web at
\mbox{http://www.batse.msfc.nasa.gov/data/pulsar}. The parameters for
Cen X-3 are based on observations with BATSE taken by Wilson et al.}.
The most significant Rayleigh power, at the X-ray period and
corresponding to a chance probability of $3\times 10^{-2}$ after
correction for period searching, was for data recorded on 1997 March 3.
One such occurrence is expectation for the number of observations made.
The 3 $\sigma$ upper limit to the gamma pulsed fraction at the
fundamental may be derived from the corresponding Rayleigh probability
(\mbox{$ = \exp{(-NR^{2})}$}), given $R$, the length of the Rayleigh
vector (\cite{kn:mardia}) and $N$, the number of events. 

The spin period history of \mbox{Centaurus X-3} (\cite{kn:bildsten})
does not allow a search for coherent pulsations involving data sets of
long duration. The data set has been split into two for coherent
analysis, comprising the observations in 1997 March and 1997 June, and
each was searched separately over a period range correponding to one
independent Fourier interval. The X-ray period of \mbox{Centaurus X-3}
during each gamma ray observation has been taken from the contemporary
BATSE results. There is no significant Rayleigh power at or close to the
expected period or half-period in either of the the two datasets. The
data sets from March and June had $N = 806$ and $N = 740$, leading to a
3$\,\sigma$ upper limit for the fraction of gamma rays which were
periodic of 9.0\% and 9.8\% respectively, for either the X-ray period or
the half-period.

Following the earlier suggestions that gamma rays may originate
elsewhere in the Centaurus X-3 system (\cite{kn:bowden},
\cite{kn:raubsmit}) we have analysed our data to search for evidence of
pulsed emission at a site other than the neutron star. In this analysis,
a matrix of points in the co-rotating frame of the system was set up and
the times of the selected events were corrected for the orbital motion
of each point. This analysis did not provide evidence of pulsed emission
at either the period or half-period from any region. 


\section{Conclusion}

Centaurus X-3 is an extremely variable source of X-rays, varying on time
scales from days to weeks. The periodic outburst reported by EGRET and
the earlier failure to detect gamma rays show that this source varies in
intensity by at least a factor of five at GeV energies over an interval
of ten months (\cite{kn:vest}). The EGRET GeV outburst, wholly pulsed at
half the X-ray period (\cite{kn:vestb}), was observed at a time of
slow-down in the X-ray period when the neutron star was not accreting
mass rapidly. Our 400 GeV unpulsed flux was observed at a time when the
BATSE data indicated a period of spin-up (March data) and a transition
between an interval of spin-down and an interval of spin-up (June data).

Our results are the first measurements of the time-averaged very high
energy (400 GeV) gamma ray flux from Centaurus X-3. The observation
method adopted by the earlier, non-imaging \v{C}erenkov telescopes could
not detect such a continuous flux --- a strong outburst of a pulsed
signal was necessary to enable a detection. These earlier detections
were confined to one night in a sample of 95 nights' data
(\cite{kn:bowden}) and one night from 59 nights' data
(\cite{kn:raubsmit}). The current detection of a lower level of
continuous emission, stable over eight nights of observation, is not in
conflict with these earlier results.

The measurements summarised in Figure \ref{fig:spectrum} suggest that
Centaurus X-3 has a continuous unpulsed emission and sporadic outbursts
of pulsed emission over a wide range of gamma ray energies. Models for
gamma ray emission predict significant absorption from $\sim 1$ GeV to
$\sim 100$ GeV due to gamma-thermal X-ray interactions in the disc
(\cite{kn:bednarek}). Future reductions in energy threshold of
\v{C}erenkov telescopes may enable the spectral variation in this
important region to be measured.

The earlier measurements of pulsed VHE gamma ray emission suggested that
the pulsed emission came from a localised area trailing the neutron star
(\cite{kn:bowden}; \cite{kn:raubsmit}). In contrast, the present
detection of unpulsed emission indicates that this radiation must
originate in an extended volume of the system.

\acknowledgements

We are grateful to the UK Particle Physics and Astronomy Research
Council for support of the project and the University of Sydney for the
lease of the Narrabri site. The Mark 6 telescope was constructed with
the assistance of the staff of the Physics Department, University of
Durham, and the efforts of Mr. P. Cottle, Mrs. E. S. Hilton and Mr. K.
Tindale are acknowledged with gratitude. We would like to thank Dr. M.
H. Finger for his helpful comments on the manuscript.

\end{document}